\documentstyle[12pt,epsfig]{article}

\newcommand{\agt}{\,\rlap{\lower 3.5 pt \hbox{$\mathchar \sim$}} \raise 1pt
 \hbox {$>$}\,}
\newcommand{\alt}{\,\rlap{\lower 3.5 pt \hbox{$\mathchar \sim$}} \raise 1pt
 \hbox {$<$}\,}

\catcode`@=11
\newcount\@tempcntc
\def\@citex[#1]#2{\if@filesw\immediate\write\@auxout{\string\citation{#2}}\fi
  \@tempcnta\z@\@tempcntb\m@ne\def\@citea{}\@cite{\@for\@citeb:=#2\do
    {\@ifundefined
       {b@\@citeb}{\@citeo\@tempcntb\m@ne\@citea\def\@citea{,}{\bf ?}\@warning
       {Citation `\@citeb' on page \thepage \space undefined}}%
    {\setbox\z@\hbox{\global\@tempcntc0\csname b@\@citeb\endcsname\relax}%
     \ifnum\@tempcntc=\z@ \@citeo\@tempcntb\m@ne
       \@citea\def\@citea{,}\hbox{\csname b@\@citeb\endcsname}%
     \else
      \advance\@tempcntb\@ne
      \ifnum\@tempcntb=\@tempcntc
      \else\advance\@tempcntb\m@ne\@citeo
      \@tempcnta\@tempcntc\@tempcntb\@tempcntc\fi\fi}}\@citeo}{#1}}
\def\@citeo{\ifnum\@tempcnta>\@tempcntb\else\@citea\def\@citea{,}%
  \ifnum\@tempcnta=\@tempcntb\the\@tempcnta\else
   {\advance\@tempcnta\@ne\ifnum\@tempcnta=\@tempcntb \else \def\@citea{--}\fi
    \advance\@tempcnta\m@ne\the\@tempcnta\@citea\the\@tempcntb}\fi\fi}
\catcode`@=12

\begin{document}

\title{
\vskip-3cm{\baselineskip14pt
\centerline{\normalsize DESY 99--085\hfill ISSN 0418--9833}
\centerline{\normalsize hep--ph/9907293\hfill}
\centerline{\normalsize July 1999\hfill}}
\vskip1.5cm
Mass Scale of New Physics in the Absence of the Higgs Boson}
\author{{\sc Bernd A. Kniehl and Alberto Sirlin}\thanks{Permanent address:
Department of Physics, New York University, 4~Washington Place, New York,
NY~10003, USA}\\
{\normalsize II. Institut f\"ur Theoretische Physik, Universit\"at Hamburg,}\\
{\normalsize Luruper Chaussee 149, 22761 Hamburg, Germany}}

\date{}

\maketitle

\thispagestyle{empty}

\begin{abstract}
We consider a hypothetical scenario in which the Higgs boson is absent, and
attempt to constrain the mass scale $\Lambda$ of the new physics that would
take its place.
Using recent measurements of $\sin^2\theta_{\mathrm{eff}}^{\mathrm{lept}}$ and
$M_W$, we show that, in a class of theories characterized by simple
conditions, the upper bound on $\Lambda$ is close to or smaller than the SM
upper bound on $M_H$, while in the complementary class $\Lambda$ is not
restricted by our considerations.
The issue of fine-tuning when $\Lambda$ is large is briefly discussed.
As a by-product of our considerations, we discuss the usefulness and important 
properties of a radiative correction, $\Delta r_{\mathrm{eff}}$, that directly
links $\sin^2\theta_{\mathrm{eff}}^{\mathrm{lept}}$ with $\alpha$, $G_\mu$,
and $M_Z$.

\medskip

\noindent
PACS numbers: 12.15.Lk, 12.20.Fv, 12.60.-i
\end{abstract}

\newpage

\section{Introduction}

For a long time, the Higgs boson and the associated Higgs mechanism have been 
viewed as one of the most intriguing pillars of the Standard Model (SM).
The consistency of this theory, its success in describing accurately a 
multitude of phenomena, and the emergence of Supersymmetry as a leading 
candidate for physics beyond the SM have given very strong support to the 
Higgs principle.
Nonetheless, the provocative questions remain:
What would happen if the Higgs boson was not found?
Is it possible to surmise something of a general nature about the new physics 
that would take its place?

In this paper, we consider a hypothetical scenario in which the dynamical 
degrees of freedom associated with the Higgs boson are absent, and attempt to 
constrain the mass scale of the unknown new physics that would take its place,
using current experimental information about two very sensitive parameters,
namely $\sin^2\theta_{\mathrm{eff}}^{\mathrm{lept}}$ and $M_W$.

\boldmath
\section{Information from $\sin^2\theta_{\mathrm{eff}}^{\mathrm{lept}}$}
\unboldmath

Our strategy is the following:
we first consider a radiative correction $\Delta r_{\mathrm{eff}}$ that
directly links the accurately measured parameter
$\sin^2\theta_{\mathrm{eff}}^{\mathrm{lept}}$, employed to analyze the data at
the $Z^0$ peak, with $\alpha$, $G_\mu$, and $M_Z$.
Specifically, we have
\begin{equation}
s_{\mathrm{eff}}^2c_{\mathrm{eff}}^2=\frac{A^2}{M_Z^2\left(1-
\Delta r_{\mathrm{eff}}\right)},
\label{e1}
\end{equation}
where $A^2=(\pi\alpha/\sqrt2G_\mu)$ and
$s_{\mathrm{eff}}^2=1-c_{\mathrm{eff}}^2$ is an  abbreviation for
$\sin^2\theta_{\mathrm{eff}}^{\mathrm{lept}}$.

An expression for $\Delta r_{\mathrm{eff}}$ can be readily obtained by
combining the relations
\begin{equation}
\hat s^2\hat c^2=\frac{A^2}{M_Z^2\left(1-\Delta\hat r\right)},
\label{e2}
\end{equation}
where $\hat s^2=1-\hat c^2=\sin^2\hat\theta_{\mathrm{w}}(M_Z)$ is the
$\overline{\mathrm{MS}}$ electroweak mixing parameter \cite{r1,r2,r3} and
\begin{equation}
s_{\mathrm{eff}}^2={\mathrm{Re}}\,\hat\kappa_{\mathrm{l}}(M_Z)\hat s^2,
\label{e3}
\end{equation}
studied in Ref.~\cite{r4}.
In Eq.~(\ref{e2}), $\Delta\hat r$ is the relevant radiative correction while, 
in Eq.~(\ref{e3}), $\hat\kappa_{\mathrm{l}}(M_Z)$ is an electroweak form
factor.
Writing
${\mathrm{Re}}\,\hat\kappa_{\mathrm{l}}(M_Z)=1+\Delta\hat\kappa_{\mathrm{l}}$,
noting that $\Delta\hat\kappa_{\mathrm{l}}$ is numerically very small
\cite{r4}, and neglecting very small $O(g^4)$ effects, one obtains
Eq.~(\ref{e1}) with the identification
\begin{equation}
\Delta r_{\mathrm{eff}}=\Delta\hat r+\Delta\hat\kappa_{\mathrm{l}}
\frac{\alpha}{\hat\alpha}\left(1-\frac{\hat s^2}{\hat c^2}\right).
\label{e4}
\end{equation}
Note that the overall coupling of
$\Delta\hat\kappa_{\mathrm{l}}(\alpha/\hat\alpha)$ is $(\alpha/\pi\hat s^2)$,
in analogy with corresponding contributions to $\Delta\hat r$ \cite{r1,r2,r3}.
In the $\overline{\mathrm{MS}}$ framework of Refs.~\cite{r1,r2,r3,r4},
$\hat s^2$ is evaluated at the scale $\mu=M_Z$ and, therefore, the same
applies to the radiative corrections $\Delta\hat r$ and
$\Delta\hat\kappa_{\mathrm{l}}$.
However, we may consider the more general situation in which $\hat s^2$, and 
therefore also $\Delta\hat r$ and $\Delta\hat\kappa_{\mathrm{l}}$, are
evaluated at a general scale $\mu$.
A very important property of Eq.~(\ref{e4}) is that, in the SM, the $\mu$ 
dependence of $\Delta\hat r$ cancels against that of
$\Delta\hat\kappa_{\mathrm{l}}(\alpha/\hat\alpha)(1-\hat s^2/\hat c^2)$, so
that $\Delta r_{\mathrm{eff}}$ is scale independent.
This also means that, if the $\overline{\mathrm{MS}}$ pole subtractions in 
$\Delta\hat r$ and
$\Delta\hat\kappa_{\mathrm{l}}(\alpha/\hat\alpha)(1-\hat s^2/\hat c^2)$ are
not implemented, their divergent parts cancel against each other.
This property, which we have verified in the SM at the one-loop level, is to 
be expected on general grounds, since $\Delta r_{\mathrm{eff}}$ is related,
via Eq.~(\ref{e1}), to physical observables.
Equation~(\ref{e1}) is analogous to the well-known expression
\begin{equation}
s^2c^2=\frac{A^2}{M_Z^2(1-\Delta r)},
\label{e5}
\end{equation}
where $s^2=1-c^2=\sin^2\theta_{\mathrm{w}}=1-M_W^2/M_Z^2$ is the electroweak
mixing parameter in the on-shell scheme of renormalization, $M_W$ and $M_Z$
are the physical masses of the intermediate bosons, and $\Delta r$ is the 
corresponding radiative correction \cite{r5}.

For reasons that will become clear later on, in the discussion of new physics,
we restrict the analysis to one-loop electroweak diagrams, but we include the
$O(\alpha\hat\alpha_{\mathrm{s}})$ corrections and the
$O(\alpha\hat\alpha_{\mathrm{s}}^2)$ contributions to $\Delta\rho$.
A convenient one-loop expression for $\Delta\hat r$, linear in the 
self-energies, is given in Eq.~(15c) of Ref.~\cite{r2}, and the corresponding 
perturbative $O(\alpha\hat\alpha_{\mathrm{s}})$ corrections can be obtained
from Ref.~\cite{r3}.
The contributions to $\Delta\hat\kappa_{\mathrm{l}}$, including QCD
corrections, are given in Ref.~\cite{r4}.
For simplicity, we neglect very small effects involving light-fermion masses.

\begin{figure}[ht]
\begin{center}
\epsfig{figure=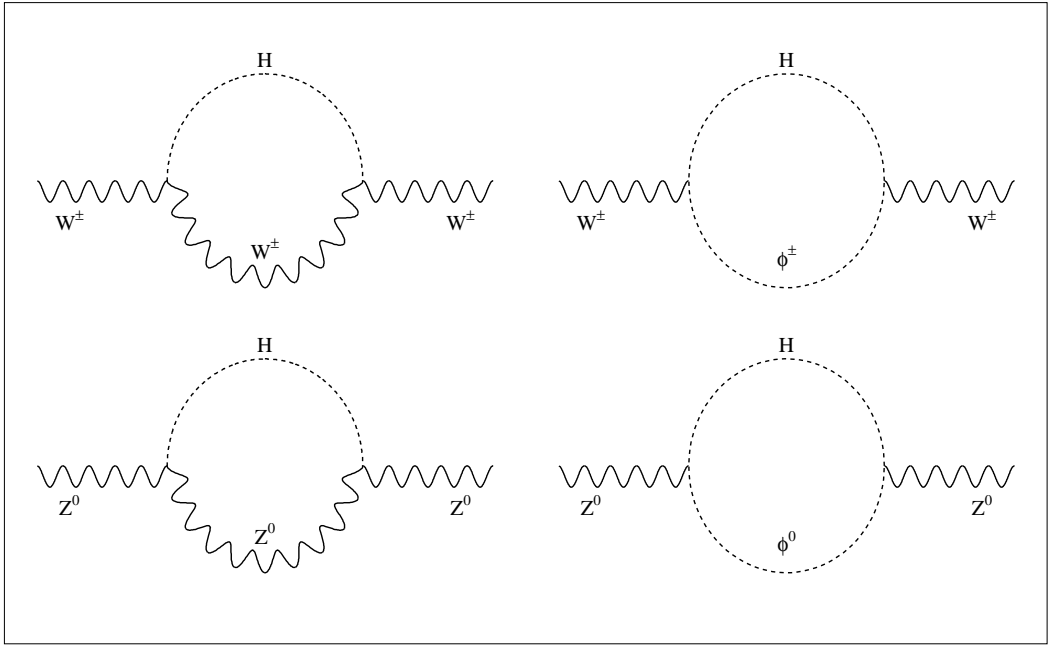,width=\textwidth}
\caption{Feynman diagrams contributing to $(\Delta r_{\mathrm{eff}})_H$ and
$\Delta r_H$ at one loop}
\label{fig:one}
\end{center}
\end{figure}

The next step in our strategy is to subtract from the SM quantity
$\Delta r_{\mathrm{eff}}$ the contribution from one-loop diagrams involving
the Higgs boson, which constitute a gauge-invariant, albeit divergent subset
\cite{r8}.
They are depicted in Fig.~\ref{fig:one}.
In the formulation of Ref.~\cite{r2}, this contribution is given by
\begin{eqnarray}
(\Delta r_{\mathrm{eff}})_H&=&\frac{\alpha}{4\pi\hat s^2\hat c^2}
\left[\left(\frac{5}{3}-\frac{3c^2}{2}\right)\left(\frac{1}{n-4}+C
+\ln\frac{M_Z}{\mu}\right)+H(\xi)-\frac{3c^2}{4}\,
\frac{\xi\ln\xi-c^2\ln c^2}{\xi-c^2}\right.\nonumber\\
&&{}+\left.\frac{19c^2}{24}+\frac{s^2}{6}\right],
\label{e6}
\end{eqnarray}
where $\xi=M_H^2/M_Z^2$, $H(\xi)$ is a function studied in Ref.~\cite{r5},
$n$ is the dimension of space-time, $C=[\gamma-\ln(4\pi)]/2$ is the
conventional constant accompanying the $1/(n-4)$ pole in dimensional
regularization, and we explicitly exhibit the divergent and $\mu$-dependent
contributions.
We note that the last four terms in Eq.~(\ref{e6}) coincide with the 
$\overline{\mathrm{MS}}$-renormalized part of $(\Delta r_{\mathrm{eff}})_H$
when the scale $\mu=M_Z$ is chosen.
Calling their contribution
$(\Delta r_{\mathrm{eff}})_H^{\overline{\mathrm{MS}}}$
and subtracting Eq.~(\ref{e6}) from $\Delta r_{\mathrm{eff}}$, we have
\begin{eqnarray}
\Delta r_{\mathrm{eff}}-(\Delta r_{\mathrm{eff}})_H
&=&-\frac{\alpha}{4\pi\hat s^2\hat c^2}
\left(\frac{5}{3}-\frac{3c^2}{2}\right)\left(\frac{1}{n-4}+C
+\ln\frac{M_Z}{\mu}\right)+\Delta r_{\mathrm{eff}}
-(\Delta r_{\mathrm{eff}})_H^{\overline{\mathrm{MS}}},\nonumber\\
&&\label{e7}
\end{eqnarray}
where
$\Delta r_{\mathrm{eff}}-(\Delta r_{\mathrm{eff}})_H^{\overline{\mathrm{MS}}}$ 
is finite and independent of $\mu$ and $M_H$.
Thus, after subtracting the Higgs-boson diagrams, we are left with a divergent
and scale-dependent expression!
Next, we conjecture that contributions from unknown new physics cancel the 
divergence and scale dependence of Eq.~(\ref{e7}).
This contribution is then of the form
\begin{equation}
(\Delta r_{\mathrm{eff}})_{\mathrm{NP}}=\frac{\alpha}{4\pi\hat s^2\hat c^2}
\left(\frac{5}{3}-\frac{3c^2}{2}\right)\left(\frac{1}{n-4}+C
+\ln\frac{M}{\mu}\right),
\label{e8}
\end{equation}
where $\ln(M/\mu)$ represents the new-physics contribution at scale $\mu$,
in the $\overline{\mathrm{MS}}$ scheme.
It is related to the scale $\Lambda$ of the new theory by
\begin{equation}
\ln\frac{M}{\mu}=\ln\frac{\Lambda}{\mu}+K,
\label{e8a}
\end{equation}
where $K=\ln(M/\Lambda)$ is the new-physics contribution at scale $\Lambda$.
Adding Eq.~(\ref{e8}) to Eq.~(\ref{e7}), we find that, in the new scenario
(NS) in which new-physics effects take the place of the Higgs-boson
contributions, the SM parameter $\Delta r_{\mathrm{eff}}$ is replaced by
\begin{equation}
(\Delta r_{\mathrm{eff}})_{\mathrm{NS}}
=\frac{\alpha}{4\pi\hat s^2\hat c^2}
\left(\frac{5}{3}-\frac{3c^2}{2}\right)\ln\frac{M}{M_Z}
+\Delta r_{\mathrm{eff}}-(\Delta r_{\mathrm{eff}})_H^{\overline{\mathrm{MS}}}.
\label{e9}
\end{equation}
The first term represents the new-physics contribution at scale $M_Z$.
As a check, we may consider the particular case in which the ``new physics" is 
provided by the SM Higgs boson.
For large $M_H$, the leading one-loop Higgs-boson contribution to
$\Delta r_{\mathrm{eff}}$ in the SM is given by
$(\alpha/4\pi\hat s^2\hat c^2)(5/3-3c^2/2)\ln(M_H/M_Z)$, so that, in the 
asymptotic limit, $M$ would then be identified with $M_H$.

Next, we constrain $\ln(M/M_Z)$ from present experimental information.
Constraints on $\Lambda$ will be discussed later on.
Inserting the current central value $s_{\mathrm{eff}}^2=0.23151\pm0.00017$ 
\cite{r9} into Eq.~(\ref{e1}), we find the experimental result
$(\Delta r_{\mathrm{eff}})_{\mathrm{exp}}=0.06053\pm0.00049$.
As $\hat s^2$ is numerically very close to $s_{\mathrm{eff}}^2$ \cite{r4}, in
the evaluation of Eq.~(\ref{e9}) we substitute
$\hat s^2\to s_{\mathrm{eff}}^2=0.23151$, $\hat c^2\to c_{\mathrm{eff}}^2$,
while for $s^2=1-c^2$ we employ 0.22223, corresponding to the current central
values, $M_W=80.419$~GeV and $M_Z=91.1871$~GeV \cite{r9}.
Using $M_t=(174.3\pm5.1)$~GeV, $\hat\alpha_{\mathrm{s}}(M_Z)=0.119\pm0.002$
\cite{r9}, and $\Delta\alpha_h^{(5)}=0.02804\pm0.00065$ \cite{r10},
Eq.~(\ref{e9})
leads to
\begin{eqnarray}
(\Delta r_{\mathrm{eff}})_{\mathrm{NS}}
&=&\frac{\alpha}{4\pi s_{\mathrm{eff}}^2c_{\mathrm{eff}}^2}
\left(\frac{5}{3}-\frac{3c^2}{2}\right)
\ln\frac{M}{M_Z}+0.06049\pm0.00065\pm0.00051\pm0.00002.\nonumber\\
&&\label{e11}
\end{eqnarray}
The quantity $\Delta\alpha_h^{(5)}$ stands for the contribution of the five 
light quark flavors $(u,d,s,c,b)$ to the renormalized photon
vacuum-polarization function evaluated at $q^2=M_Z^2$.
Later on, we illustrate the corresponding results for the value
$\Delta\alpha_h^{(5)}=0.02770\pm0.00016$ \cite{r11}, determined by applying
perturbative QCD down to energies of a few GeV.
The errors in Eq.~(\ref{e11}) are induced by those in $\Delta\alpha_h^{(5)}$,
$M_t$, and $\hat\alpha_{\mathrm{s}}(M_Z)$, respectively.
The last two are found by independently varying $M_t$ and
$\hat\alpha_{\mathrm{s}}(M_Z)$ in the evaluation of Eq.~(\ref{e9}).
Comparing Eq.~(\ref{e11}) with $(\Delta r_{\mathrm{eff}})_{\mathrm{exp}}$ and
combining the errors in quadrature, we obtain
\begin{equation}
\ln\frac{M}{M_Z}=0.028\pm0.586,
\label{e12}
\end{equation}
corresponding to the central value $M_{\mathrm{c}}=94$~GeV and an upper bound
$M_{95}=245$~GeV at the 95\% confidence level (CL).
We note that we have employed QCD corrections to $\Delta\rho$ in the 
conventional $M_t$ approach \cite{r12}, which, in calculations restricted to 
one-loop electroweak amplitudes, leads to larger values of
$\ln(M/M_Z)$ than optimized methods \cite{r13}.
Thus, as far as uncertainties in the QCD corrections are concerned, we take 
Eq.~(\ref{e12}) to be a rather conservative estimate.

\boldmath
\section{Information from $M_W$}
\unboldmath

Turning our attention to $\Delta r$, we employ Eq.~(21) of Ref.~\cite{r2} with 
the approximation $(c^2/s^2)\Delta\hat\rho(1-\Delta\hat r_{\mathrm{w}})\approx
(c^2/s^2)\Delta\hat\rho(\alpha/\hat\alpha)$, where $\Delta\hat\rho$ and
$\Delta\hat r_{\mathrm{w}}$ are radiative corrections defined in that work.
In this way, we obtain an expression for $\Delta r$ that depends linearly on 
the self-energies.
In the SM, this approximation neglects two-loop effects of order 
$2\times10^{-4}$.
The SM Higgs-boson contribution to $\Delta r$ in the SM is given by
\begin{eqnarray}
\Delta r_H&=&\frac{\alpha}{4\pi\hat s^2}
\left[\frac{11}{6}\left(\frac{1}{n-4}+C+\ln\frac{M_Z}{\mu}\right)
+\frac{H(\xi)-H(\xi/c^2)(1-2s^2)}{s^2}\right.\nonumber\\
&&{}-\left.\frac{3\xi}{4(\xi-c^2)}\ln\frac{\xi}{c^2}
+\frac{23}{24}+\frac{\ln c^2}{6}\left(\frac{1}{2}-\frac{5c^2}{s^2}
\right)\right].
\label{e13}
\end{eqnarray}
The relevant Feynman diagrams are depicted in Fig.~\ref{fig:one}.
The expression analogous to Eq.~(\ref{e9}) becomes
\begin{equation}
\Delta r_{\mathrm{NS}}=\frac{11\alpha}{24\pi\hat s^2}\ln\frac{M^\prime}{M_Z}
+\Delta r-\Delta r_H^{\overline{\mathrm{MS}}},
\label{e14}
\end{equation}
where the first term represents the new-physics contribution at scale $M_Z$
and $\Delta r$ is the SM correction.
Employing the same input parameters as in Eq.~(\ref{e11}), Eq.~(\ref{e14}) 
leads to
\begin{equation}
\Delta r_{\mathrm{NS}}=\frac{11\alpha}{24\pi s_{\mathrm{eff}}^2}
\ln\frac{M^\prime}{M_Z}+0.03355\pm0.00065\pm0.00191\pm0.00007.
\label{e15}
\end{equation}
Inserting $M_W=(80.419\pm0.038)$~GeV and $M_Z=(91.1871\pm0.0021)$~GeV
\cite{r9} into Eq.~(\ref{e5}), we obtain the experimental value
$\Delta r_{\mathrm{exp}}=0.03298\pm0.00230$.
Comparison of Eq.~(\ref{e15}) with $\Delta r_{\mathrm{exp}}$ leads to
\begin{equation}
\ln\frac{M^\prime}{M_Z}=-0.124\pm0.666,
\label{e17}
\end{equation}
corresponding to $M_{\mathrm{c}}^\prime=81$~GeV and $M_{95}^\prime=240$~GeV.
We note that $M_{\mathrm{c}}^\prime$ and $M_{95}^\prime$ are very close to
$M_{\mathrm{c}}$ and $M_{95}$, respectively.

\section{Discussion and Conclusions}

It is interesting to compare the above results for $M$ and $M^\prime$ with SM
estimates of $M_H$.
Using the same input parameters leading to Eq.~(\ref{e12}), we apply Eq.~(3) of 
Ref.~\cite{r14} in the $\overline{\mathrm{MS}}$ scheme discussed in that paper
to extract $\ln(M_H/M_Z)$ from $s_{\mathrm{eff}}^2$ in the SM, with the result
\begin{equation}
\ln\frac{M_H}{M_Z}=0.016\pm0.629,
\label{e18}
\end{equation}
which corresponds to a central value $M_H^{\mathrm{c}}=93$~GeV and a 95\% CL
upper bound $M_H^{95}=260$~GeV.
If, instead, $M_W$ is employed as the basic input, using Eq.~(4) of 
Ref.~\cite{r14} in the $\overline{\mathrm{MS}}$ scheme, we find in the SM
\begin{equation}
\ln\frac{M_H}{M_Z}=-0.728{+0.972\atop-1.596},
\label{e19}
\end{equation}
$M_H^{\mathrm{c}}=44$~GeV and $M_H^{95}=193$~GeV.
We see that Eq.~(\ref{e18}), the SM determination of $M_H$ derived from 
$s_{\mathrm{eff}}^2$, is very close to Eq.~(\ref{e12}), the $M$ estimate from
the same source.
On the other hand, the SM estimate of $M_H$ derived from $M_W$ 
[Eq.~(\ref{e19})] is somewhat more restrictive than the corresponding
$M^\prime$ determination [Eq.~(\ref{e17})], with $M_H^{\mathrm{c}}$ and
$M_H^{95}$ lower by about 37~GeV and 47~GeV, respectively.
Nonetheless, the overall picture that emerges is that the $M$ and $M^\prime$
estimates are very close and also quite similar to the SM determinations of
$M_H$.

The same pattern is apparent when one employs the ``theory-driven" 
determination of $\Delta\alpha_h^{(5)}$, yielding
$\Delta\alpha_h^{(5)}=0.02770\pm0.00016$ \cite{r11}.
In this case, using $s_{\mathrm{eff}}^2$ as input, we find
$M_{\mathrm{c}}=116$~GeV and $M_{95}=239$~GeV, while the SM values for $M_H$
are $M_H^{\mathrm{c}}=116$~GeV and $M_H^{95}=249$~GeV.
If, instead, $M_W$ is used as input, we obtain $M_{\mathrm{c}}^\prime=87$~GeV
and $M_{95}^\prime=252$~GeV, while the SM values are $M_H^{\mathrm{c}}=51$~GeV
and $M_H^{95}=205$~GeV.
We also note that the central values obtained with this determination of
$\Delta\alpha_h^{(5)}$ are somewhat larger than in the case of
$\Delta\alpha_h^{(5)}=0.02804\pm0.00065$ \cite{r10}, but the 95\% CL upper
bounds are generally close.

The closeness of $M$, $M^\prime$, and the corresponding $M_H$ estimates in the
SM is perhaps surprising.
We have already pointed out, in the discussion after Eq.~(\ref{e9}), that at 
the one-loop level one expects $M$ and $M^\prime$ to approach asymptotically
the SM estimate for $M_H$.
However, in the range 100~GeV${}\alt M_H\alt{}$300~GeV, we are far away from 
the asymptotic domain.
Furthermore, there are important differences between our approach to evaluate 
$M$ and $M^\prime$, and the SM calculation of $M_H$.
In the latter, one considers the detailed $M_H$ dependence of the one-loop 
corrections, and, moreover, current studies include complicated two-loop 
$M_H$-dependent effects of $O(\alpha^2M_t^4/M_W^4)$ and
$O(\alpha^2M_t^2/M_W^2)$ \cite{r16}.
It is known that, in the SM, these two-loop effects decrease significantly the 
derived value for $M_H$ and its upper bound.
In the hypothetical scenario we consider in this paper, in which the degrees 
of freedom associated with the Higgs boson are absent, it is not possible to 
incorporate such $M_H$-dependent effects, and, for that reason, we have 
restricted ourselves to one-loop electroweak diagrams.
Nonetheless, the results of the two approaches are very similar.
In particular, it is quite remarkable that the simple formula of 
Eq.~(\ref{e11}) with the replacement $M\to M_H$ reproduces very accurately the
SM calculation for $M_H\approx100$~GeV and remains a reasonable approximation
for larger values of $M_H$.
For instance, for $M_H=100$, 300, 600~GeV, the differences between 
Eq.~(\ref{e11}) and the SM calculations of Ref.~\cite{r14} amount to
$-1\times10^{-5}$, $1.6\times10^{-4}$, and $2.7\times10^{-4}$ (corresponding
to differences in $s_{\mathrm{eff}}^2$ of less than $1\times10^{-5}$,
$6\times10^{-5}$, and $1\times10^{-4}$), respectively.
Over the same $M_H$ range, Eq.~(\ref{e15}) leads to differences in $M_W$ of 23 
to 10~MeV, relative to Ref.~\cite{r14}.

We now discuss the implications of these results for the scale $\Lambda$ of
the unknown, alternative theory.
Following Eq.~(\ref{e8a}), we write
\begin{equation}
\ln\frac{M}{M_Z}=\ln\frac{\Lambda}{M_Z}+K
\label{e20}
\end{equation}
in the case of $\Delta r_{\mathrm{eff}}$ [Eqs.~(\ref{e9})--(\ref{e12})], and
\begin{equation}
\ln\frac{M^\prime}{M_Z}=\ln\frac{\Lambda}{M_Z}+K^\prime
\label{e21}
\end{equation}
in the case of $\Delta r$ [Eqs.~(\ref{e14})--(\ref{e17})].
As the alternative theory is not specified, $K$ and $K^\prime$ are unknown
constants.
Equations~(\ref{e20}) and (\ref{e21}) permit us to classify all alternative
theories into two broad categories:
(1) theories in which either $K$ or $K^\prime$ (or both) are positive;
(2) theories in which both $K$ and $K^\prime$ are negative.
We see that theories of the first class are bounded by one of the two
equations.
If $K$ is positive, at 95\% CL we have $\Lambda\alt245$~GeV [Eq.~(\ref{e12})].
If $K^\prime$ is positive, $\Lambda\alt240$~GeV [Eq.~(\ref{e17})],
independently of the precise value of $|K|$.
We note that these limits are approximately of the same magnitude as those
currently derived for $M_H$ in the SM.
It is worth noting that $K^\prime-K=-0.152\pm0.887$, so that one can entertain
the possibility that both $K$ and $K^\prime$ are simultaneously small, which
may occur in a subclass among weakly interacting theories.
If $|K|,|K^\prime|\alt0.1$, for instance, the $\Lambda$ bounds would be quite
close to the SM bounds on $M_H$.

Theories of the second class evade our bounds and may correspond to large
values of $\Lambda$, as in the scenario recently discussed in Ref.~\cite{r17}.
If $\Lambda=1$~TeV, for instance, we have $K=-2.367\pm0.586$
[Eqs.~(\ref{e12}) and (\ref{e20})] and $K^\prime=-2.519\pm0.666$
[Eqs.~(\ref{e17}) and (\ref{e21})].
Thus, for $\Lambda=1$~TeV, in order to reproduce our central values, a
fine-tuning of the parameters $K$ and $K^\prime$ is necessary.
However, if $K$ and $K^\prime$ differ from their central values by about one
sigma, one may argue that no excessive fine-tuning is necessary.
On the other hand, even at the one-sigma level, a substantial cancellation
between the logarithmic and constant terms is required in order to accomodate
our results, an observation that may become sharper in the future as errors
decrease.

It is important to emphasize that in our formulation we do not assume {\it ab
initio} that the scale $\Lambda$ of the new physics is large relative to
$M_Z$.
In fact, our procedure has been to focus on the radiative corrections
associated with two very precise observables, subtract exactly the Higgs-boson
contribution, and then restrict $\Lambda$ from accurate experiments.
This leads to our conclusion that $K>0$ and/or $K^\prime>0$ is a sufficient
condition for $\Lambda$ to have low 95\% CL upper bounds.
The usefulness of this condition stems from the fact that, if satisfied, one
can draw immediately the conclusion that $\Lambda$ is sharply restricted.
In some alternative formulations, such as those based on the $S$, $T$, and $U$
parameters, one assumes {\it ab initio} that $\Lambda$ is large relative to
$M_Z$.
Although such formulations permit to discuss the large-$\Lambda$ case, they
are clearly not suitable to draw conclusions or derive sufficiency conditions
in the case when $\Lambda$ is close to the electroweak scale.

We also note that the sharp constraints derived above for theories of the 
first class are based on the smallness of $\ln(M/M_Z)$ in Eq.~(\ref{e12}) and
$\ln(M^\prime/M_Z)$ in Eq.~(\ref{e17}).
In turn, this is related to the very curious fact that the SM Higgs-boson
contributions to $(\Delta r_{\mathrm{eff}})_H^{\overline{\mathrm{MS}}}$ and
$\Delta r_H^{\overline{\mathrm{MS}}}$ are highly suppressed for current
experimental inputs.
In order to show this, it is sufficient to point out that
$(\alpha/4\pi s_{\mathrm{eff}}^2c_{\mathrm{eff}}^2)(5/3-3c^2/2)\ln(M/M_Z)$ in
Eq.~(\ref{e11}) and
$(11\alpha/24\pi s_{\mathrm{eff}}^2)\ln(M^\prime/M_Z)$ in Eq.~(\ref{e15})
replace the full one-loop Higgs-boson contributions, as well as the
$M_H$-dependent effects of $O(\alpha^2M_t^4/M_W^4)$ and
$O(\alpha^2M_t^2/M_W^2)$, currently incorporated in the calculation of these
basic corrections \cite{r14,r16}.
Using the values of $\ln(M/M_Z)$ in Eq.~(\ref{e12}) and $\ln(M^\prime/M_Z)$ in
Eq.~(\ref{e17}), we find that these contributions amount to only
$(0.46\pm9.57)\times10^{-4}$ and $(-5.69\pm30.61)\times10^{-4}$, respectively.
As the typical order of magnitude of electroweak corrections is
$\alpha/(\pi s_{\mathrm{eff}}^2)\approx0.01$, we conclude that the SM
Higgs-boson contributions to
$(\Delta r_{\mathrm{eff}})_H^{\overline{\mathrm{MS}}}$ and
$\Delta r_H^{\overline{\mathrm{MS}}}$ are, indeed, highly suppressed.
This corresponds to the small values of $\ln(M/M_Z)$ and
$\ln(M^\prime/M_Z)$ found above, and leads to the sharp constraints on 
theories of the first class.
As an illustrative counter-example, if the SM Higgs-boson contributions to
these corrections were 0.5\% for current experimental inputs, and the errors
were the same, we would obtain $\ln(M/M_Z)=3.064\pm0.586$ and
$\ln(M^\prime/M_Z)=1.087\pm0.666$, corresponding to $M_{95}=5.10$~TeV and
$M_{95}^\prime=806$~GeV, and the constraints on theories of the first class
would be irretrievably lost.

The simplest mechanism that gives positive contributions to $K$ and $K^\prime$
is a heavy, ordinary, degenerate fermion doublet.
It contributes $2/11$ to $K^\prime$ \cite{r18} and, to very good
approximation, $1/3$ to $K$.
Theories that contain several degenerate doublets, such as the simplest
technicolor models, are likely to belong to the first class.
Instead, heavy non-degenerate doublets with sufficiently large mass splitting
contribute negatively to $K$ and $K^\prime$ and are likely to lead to
second-class theories.

In this paper, we have considered the constraints derived from the two most
sensitive measurements, to wit those of
$\sin^2\theta_{\mathrm{eff}}^{\mathrm{lept}}$ and $M_W$.
Additional information can be obtained from other observables, such as the 
$Z$-boson width.
However, the corresponding effective logarithm is not expected to be as
tightly constrained as those in Eqs.~(\ref{e12}) and (\ref{e17}).

In summary, there are very powerful arguments to expect the discovery of the
Higgs boson.
In the unlikely event that this fundamental particle is not found, one expects
the emergence of new physics.
We have seen that in a class of alternative theories, characterized by a
simple condition, the scale $\Lambda$ is bounded by limits that are roughly of
the same magnitude as those currently derived for $M_H$ in the SM, while the
complementary class is not restricted by our considerations.
As a by-product, we have emphasized the usefulness of a basic electroweak
correction, $\Delta r_{\mathrm{eff}}$, that directly links
$s_{\mathrm{eff}}^2$, $\alpha$, $G_\mu$, and $M_Z$.

\vspace{1cm}
\noindent
{\it Acknowledgements.}
The authors are indebted to A. Masiero, who raised the question of how the
estimates of $\Lambda$ and $M_H$ may be related.
A.S. thanks G. Degrassi and P. Gambino for very useful communications, the
members of the $2^{\mathrm{nd}}$ Institute for Theoretical Physics of Hamburg
University for their warm hospitality, and the Alexander von Humboldt
Foundation for its kind support.
The research of B.A.K. was supported in part by the Deutsche 
Forschungsgemeinschaft under Contract No.\ KN~365/1-1, by the
Bundesministerium f\"ur Bildung und Forschung under Contract No.\ 05~HT9GUA~3,
and by the European Commission through the Research Training Network
{\it Quantum Chromodynamics and the Deep Structure of Elementary Particles}
under Contract No.\ ERBFMRX-CT98-0194.
The research of A.S. was supported in part by National Science Foundation 
through Grant No.\ PHY-9722083.

\newpage

\end{document}